\documentclass[cits]{PoS}\usepackage{amsmath,cite,color}

\title{Constraints from the $1/N_{\rm c}$ Expansion on Properties
of Exotic Tetraquark Mesons}\ShortTitle{Constraints from the
$1/N_{\rm c}$ Expansion on Properties of Exotic Tetraquark Mesons}
\author{\speaker{Wolfgang Lucha}\\Institute for High Energy
Physics, Austrian Academy of Sciences, Nikolsdorfergasse 18,
A-1050 Vienna, Austria\\E-mail: \email{Wolfgang.Lucha@oeaw.ac.at}}
\author{Dmitri Melikhov\\Institute for High Energy
Physics, Austrian Academy of Sciences, Nikolsdorfergasse 18,
A-1050 Vienna, Austria, and\\ D.~V.~Skobeltsyn Institute of
Nuclear Physics, M.~V.~Lomonosov Moscow State University, 119991,
Moscow, Russia, and\\ Faculty of Physics, University of Vienna,
Boltzmanngasse 5, A-1090 Vienna, Austria\\E-mail:
\email{dmitri\_melikhov@gmx.de}}
\author{Hagop Sazdjian\\Institut de Physique Nucl\'eaire,
CNRS-IN2P3, Universit\'e Paris-Sud, Universit\'e Paris-Saclay,
91406 Orsay Cedex, France\\E-mail: \email{sazdjian@ipno.in2p3.fr}}

\abstract{Scrutinizing the scattering of ordinary mesons in the
limiting case of the number of colour degrees of freedom $N_{\rm
c}$ of quantum chromodynamics approaching infinity, we formulate
Feynman-diagram selection criteria and from these deduce rigorous
self-consistency conditions for the manifestation of a tetraquark,
a two-quark--two-antiquark bound state, as a pole in the
corresponding amplitudes. Our constraints bear rather far-reaching
consequences: In particular, all flavour-exotic tetraquarks,
composed of four (anti)quarks of disparate flavour, must come in,
at least, two variants differing in (and thus readily identifiable
by) the large-$N_{\rm c}$ behaviour of their couplings to two
ordinary~mesons. Quite generally, irrespective of their flavour
composition, all tetraquarks prove to be narrow. Their decay rates
behave, for large $N_{\rm c},$ like $1/N_{\rm c}^2$ and thus
decrease faster than those of ordinary mesons.}

\FullConference{XVII International Conference on Hadron
Spectroscopy and Structure --- Hadron2017\\25--29 September,
2017\\University of Salamanca, Salamanca, Spain}\begin{document}

\section{Criteria for the Potential Existence of Polyquark Hadron
States in Large-$N_{\rm c}$ QCD}Large-$N_{\rm c}$ QCD \cite{GH} is
an element of a class of quantum field theories obtained as
generalizations of quantum chromodynamics by allowing $N_{\rm c}$,
the number of the colour degrees of freedom of QCD, to differ from
$N_{\rm c}=3$. It is defined by considering the limit of $N_{\rm
c}$ increasing beyond bounds, $N_{\rm c}\to\infty$, with
correlative decrease of the strong coupling $\alpha_{\rm s}\equiv
g_{\rm s}^2/4\pi$, $\alpha_{\rm s}\propto1/N_{\rm c}$, and all
quarks transforming according to the $N_{\rm c}$-dimensional
fundamental representation of its underlying gauge~group ${\rm
SU}(N_{\rm c})$.

Polyquarks are exotic hadrons that cannot be interpreted as just
quark--antiquark or three-quark bound states; among these,
tetraquarks are considered to be essentially composed of two
quarks and two antiquarks. Within the framework of large-$N_{\rm
c}$ QCD, only recently, after going~through a kind of chequered
history, the theoretical opinion about existence and observability
of tetraquarks started to consolidate: At the leading order of
their $1/N_{\rm c}$ expansion, all QCD Green functions are
saturated by free mesons as intermediate states \cite{EW}.
Tetraquarks may be formed only at $N_{\rm c}$-subleading orders
\cite{SC}; unlike ordinary mesons, tetraquarks were expected not
to survive the limit $N_{\rm c}\to\infty$ as stable mesons.
Mistakenly, the latter has been regarded as an argument for the
tetraquarks' non-existence in nature.

From the point of view of experimental evidence, however, the
relevant feature of tetraquarks~is the large-$N_{\rm c}$ behaviour
of their decay rates \cite{SW}, which must not grow with $N_{\rm
c}$: observable tetraquarks should be narrow states. Tetraquarks
with all quark flavours different exhibit decay rates decreasing
like $1/N_{\rm c}^2$ \cite{X}; so they are even narrower than
ordinary mesons, whose decay rates are of order~$1/N_{\rm c}$.

In that situation, we embarked on a systematic analysis
\cite{TQ1,TQ2} of the necessary conditions for the appearance of
tetraquarks, $T$, with masses $m_T$ assumed to remain finite for
$N_{\rm c}\to\infty$, in~form~of~poles corresponding to bound
states of four (anti)quarks, with masses $m_1$, $m_2$, $m_3$, and
$m_4$, in the $s$ channel of scattering processes of two ordinary
mesons into two ordinary mesons. To this end, neglecting~all
reference to spin or parity irrelevant in the present context, we
study four-point correlation functions of operators
$j_{ij}\equiv\bar q_i\,q_j$, bilinear in the quarks and
interpolating the scattered mesons, $M_{ij}$, consisting of
antiquark $\bar q_i$ and quark $q_j$ (of flavour $i,j=1,2,3,4$),
that is, having~nonvanishing matrix elements between vacuum and
meson state, $\langle 0|j_{ij}|M_{ij}\rangle\equiv
f_{M_{ij}}\ne0$, which rise for large $N_{\rm c}$ like
$f_{M_{ij}}\propto\sqrt{N_{\rm c}}$ \cite{EW}.

With respect to the contributions to the expansion of the
corresponding scattering amplitudes~in powers of $1/N_{\rm c}$ and
$\alpha_{\rm s}$, our foremost task is to provide the
characterization of the ``tetraquark-phile'' Feynman diagrams
(with subscripts T indicating their correlator contributions) that
might develop a tetraquark pole at $s=m_T^2$ by two basic
criteria, expressed in terms of meson momenta $p_1$ and~$p_2$
\cite{TQ1}:\begin{enumerate}\item The Feynman diagram should
depend in a nonpolynomial way on the variable
$s\equiv(p_1+p_2)^2$.\item The Feynman diagram should admit
adequate intermediate four-quark states with branch cuts starting
at branch points $s=(m_1+m_2+m_3+m_4)^2$. The potential presence
or absence of such an intermediate-state threshold may be clearly
decided by means of the Landau equations\cite{LDL}.\end{enumerate}

\section{Rigorous Self-Consistency Conditions from Large-$N_{\rm
c}$ Ordinary-Meson Scattering}In order to carve out
characteristics of tetraquarks for large $N_{\rm c}$, we identify
our tetraquark-phile Feynman diagrams and derive the
tetraquark--two-ordinary-meson amplitudes $A(T\leftrightarrow
M\,M)$ that fix the tetraquarks' decay widths $\Gamma(T)$. Since
we take into account all conceivable scattering processes, we
expect to encounter ``flavour-preserving'' ones, where the flavour
composition of initial and final ordinary mesons is identical, and
``flavour-rearranging'' ones, proceeding by a reshuffling of
quarks.

\begin{figure}[b]\begin{center}\begin{tabular}{c}
\includegraphics[scale=.30115]{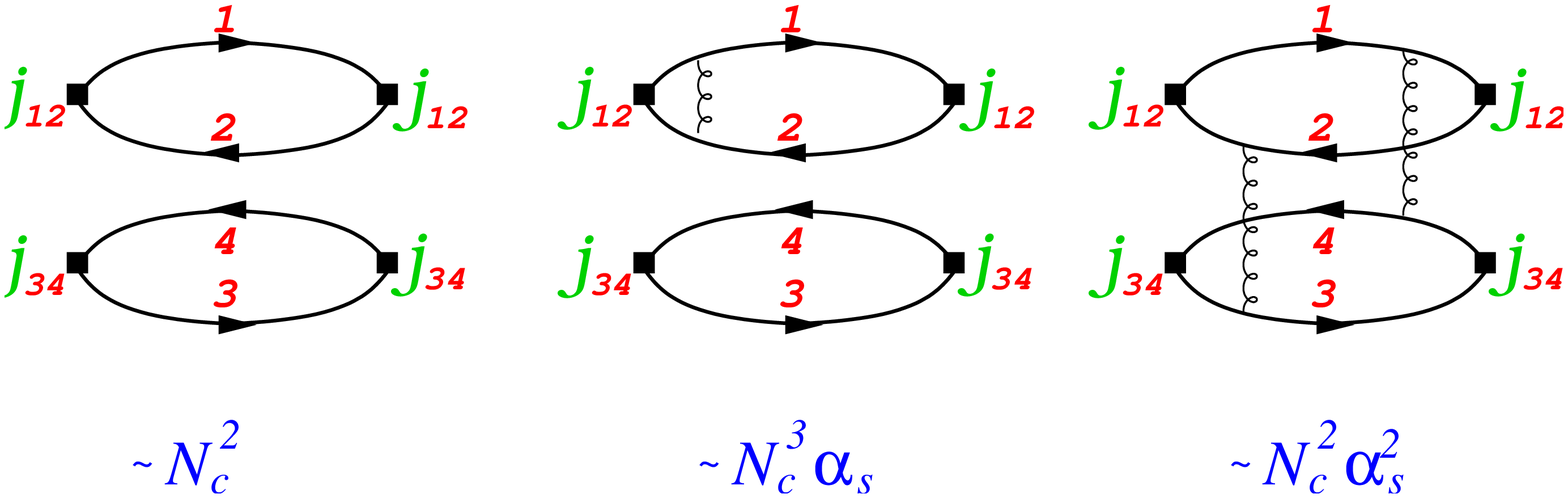}\\(a)\\[2ex]
\includegraphics[scale=.30115]{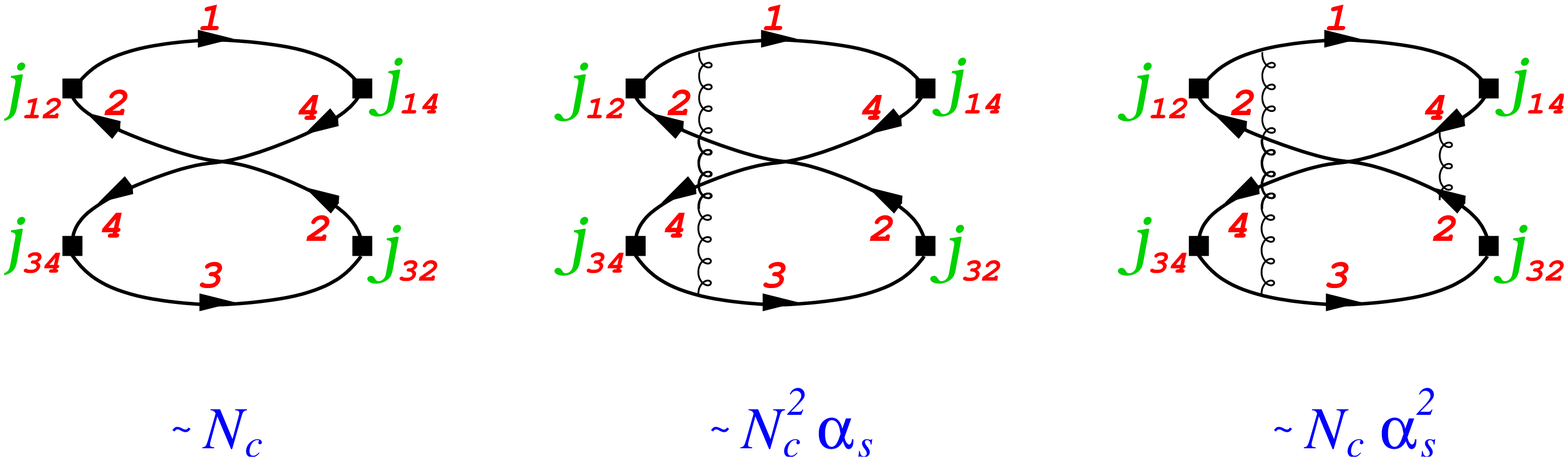}\\(b)\end{tabular}
\caption{Flavour-preserving, $\langle j^\dag_{12}\,j^\dag_{34}\,
j_{12}\,j_{34}\rangle$ (a), or -rearranging, $\langle j^\dag_{14}
\,j^\dag_{32}\,j_{12}\,j_{34}\rangle$ (b), four-point
correlators.}\label{F:E}\end{center}\end{figure}

\subsection{Flavour-exotic tetraquarks: bound states of quarks
involving four different flavour types}For genuinely
flavour-exotic tetraquarks, the rightmost plots in Fig.~\ref{F:E}
reveal that the $N_{\rm c}$-leading terms in tetraquark-phile
flavour-preserving and flavour-reshuffling four-point correlation
functions exhibit a different large-$N_{\rm c}$ behaviour. In this
case, the resulting constraints cannot be fulfilled~under the
premise of the existence of a single tetraquark; rather, they
require pole contributions of, at least, two tetraquark states,
called $T_A$ and $T_B$, differing in their large-$N_{\rm c}$
couplings to two ordinary mesons. Expressed in terms of
tetraquark--two-meson transition amplitudes $A$, generic decay
constants $f_M$ of ordinary mesons, and propagator poles at
$p^2=m^2_{T_{A,B}}$, the pole contributions then read,
symbolically,\begin{align*}\langle
j^\dag_{12}\,j^\dag_{34}\,j_{12}\,j_{34}\rangle_{\rm T}&=f_M^4
\left(\frac{|A(M_{12}\,M_{34}\leftrightarrow T_A)|^2}
{p^2-m^2_{T_A}}+\frac{|A(M_{12}\,M_{34}\leftrightarrow
T_B)|^2}{p^2-m^2_{T_B}}\right)+\cdots=O(N_{\rm c}^0)\
,\\[1ex]\langle
j^\dag_{14}\,j^\dag_{32}\,j_{14}\,j_{32}\rangle_{\rm T}&=f_M^4
\left(\frac{|A(M_{14}\,M_{32}\leftrightarrow T_A)|^2}
{p^2-m^2_{T_A}}+\frac{|A(M_{14}\,M_{32}\leftrightarrow
T_B)|^2}{p^2-m^2_{T_B}}\right)+\cdots=O(N_{\rm c}^0)\
,\\[1ex]\langle
j^\dag_{14}\,j^\dag_{32}\,j_{12}\,j_{34}\rangle_{\rm T}&=f_M^4
\left(\frac{A(M_{12}\,M_{34}\leftrightarrow T_A)\,
A(T_A\leftrightarrow M_{14}\,M_{32})}{p^2-m^2_{T_A}}\right.\\&
\hspace{4.66ex}+\left.\frac{A(M_{12}\,M_{34}\leftrightarrow T_B)\,
A(T_B\leftrightarrow M_{14}\,M_{32})}{p^2-m^2_{T_B}}\right)+\cdots
=O(N_{\rm c}^{-1})\ .\end{align*}The amplitudes of leading order
in $N_{\rm c}$ then govern the total decay rates of $T_A$ and
$T_B$; these, due to an inherent flavour symmetry of the above,
prove to show a parametrically identical dependence~on~$N_{\rm
c}$:\begin{align*}A(T_A\leftrightarrow M_{12}\,M_{34})=O(N_{\rm
c}^{-1})\ ,\qquad A(T_A\leftrightarrow M_{14}\,M_{32})=O(N_{\rm
c}^{-2})\qquad&\Longrightarrow\qquad\Gamma(T_A)=O(N_{\rm c}^{-2})\
,\\[1ex]A(T_B\leftrightarrow M_{12}\,M_{34})=O(N_{\rm c}^{-2})\
,\qquad A(T_B\leftrightarrow M_{14}\,M_{32})=O(N_{\rm
c}^{-1})\qquad&\Longrightarrow\qquad\Gamma(T_B)=O(N_{\rm c}^{-2})\
.\end{align*}Of course, as bound states composed of the same set
of constituents, $T_A$ and $T_B$ will undergo mixing.

\subsection{Flavour-cryptoexotic tetraquarks: four-quark states
exhibiting merely two open flavours}
\begin{figure}[t]\begin{center}\begin{tabular}{cc}
\includegraphics[scale=.30115]{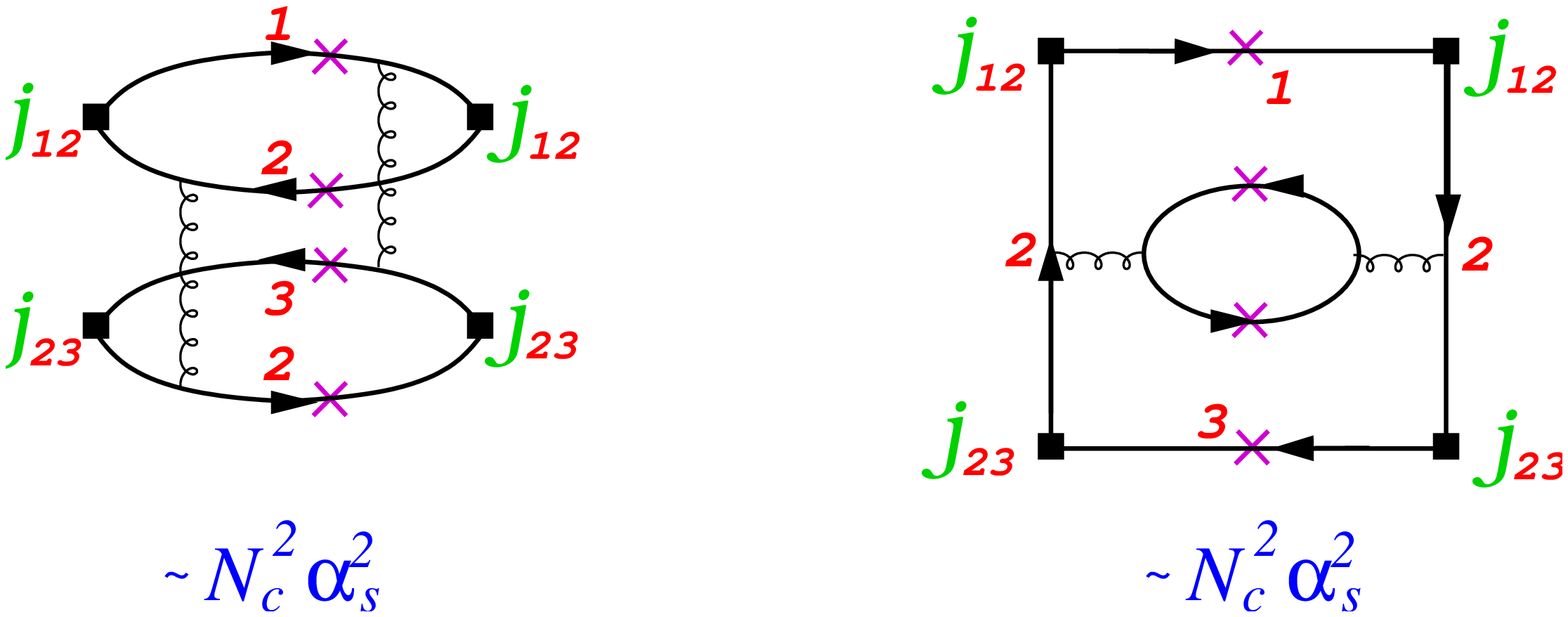}\qquad&\qquad
\includegraphics[scale=.30115]{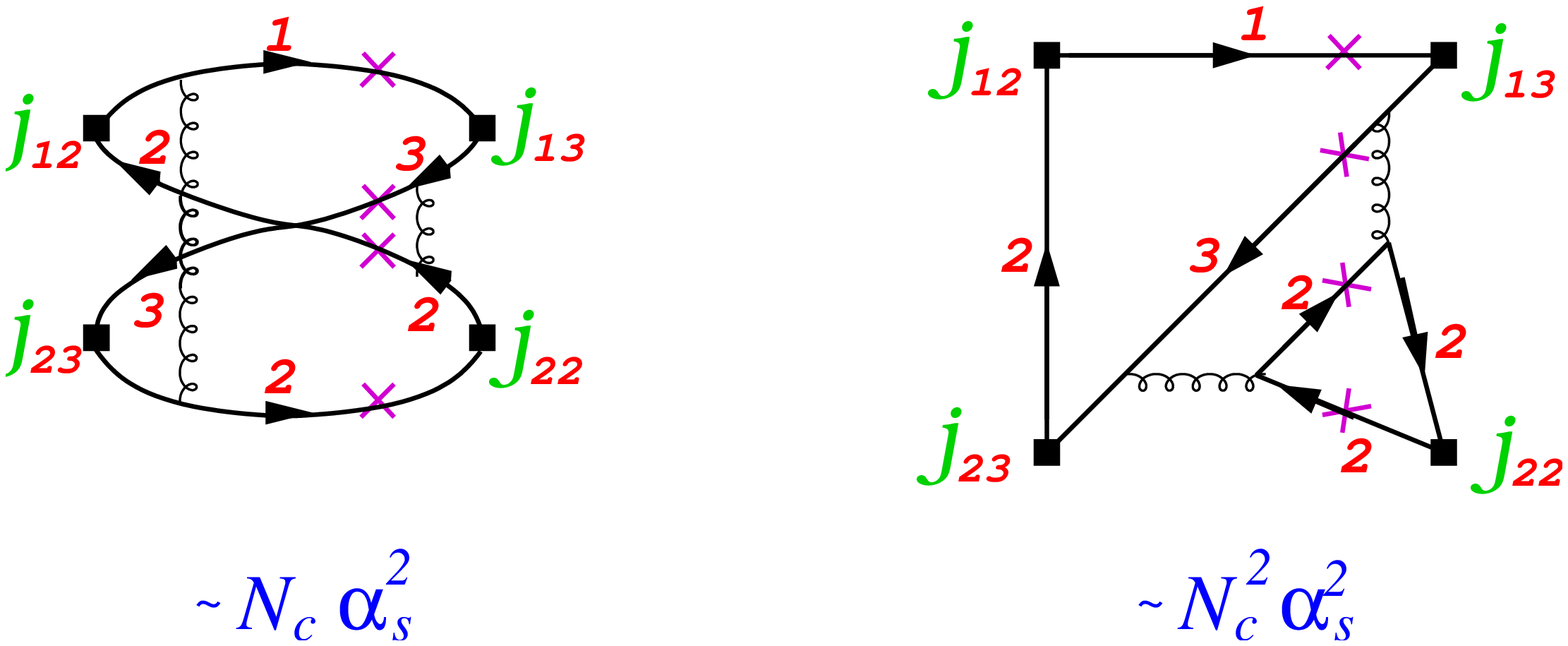}\\(a)&(b)
\end{tabular}\caption{Flavour-preserving, $\langle j^\dag_{12}\,
j^\dag_{23}\,j_{12}\,j_{23}\rangle$ (a), or -rearranging, $\langle
j^\dag_{13}\,j^\dag_{22}\,j_{12}\,j_{23}\rangle$ (b), four-point
correlators; the four purple crosses at the quark propagators
indicate the constituents of the intermediate-state tetraquarks.}
\label{F:C}\end{center}\end{figure}

For a quark--antiquark pair
with same flavour different from the other two, the relations
(Fig.~\ref{F:C})$$\langle
j^\dag_{12}\,j^\dag_{23}\,j_{12}\,j_{23}\rangle_{\rm T}=O(N_{\rm
c}^0)\ ,\qquad\langle
j^\dag_{13}\,j^\dag_{22}\,j_{13}\,j_{22}\rangle_{\rm T}=O(N_{\rm
c}^0)\ ,\qquad\langle
j^\dag_{13}\,j^\dag_{22}\,j_{12}\,j_{23}\rangle_{\rm T}=O(N_{\rm
c}^0)$$are satisfied by a single tetraquark with two-meson
couplings implying again a narrow decay width:$$A(T\leftrightarrow
M_{12}\,M_{23})=O(N_{\rm c}^{-1})\ ,\qquad A(T\leftrightarrow
M_{13}\,M_{22})=O(N_{\rm c}^{-1})\qquad\Longrightarrow\qquad
\Gamma(T)=O(N_{\rm c}^{-2})\ .$$Due to the same total flavour,
such tetraquark may mix with an ordinary meson $M_{13}$. The
pole~terms $$\langle
j^\dag_{12}\,j^\dag_{23}\,j_{12}\,j_{23}\rangle_{\rm T}
=f_M^4\left(\frac{A(M_{12}\,M_{23}\to T)}{p^2-m^2_T}\,g_{TM_{13}}
\,\frac{A(M_{13}\to M_{12}\,M_{23})}{p^2-m^2_{M_{13}}}\right)
+\cdots =O(N_{\rm c}^0)$$then restrict the large-$N_{\rm c}$
behaviour of their associated mixing strength $g_{TM_{13}}$ to
$g_{TM_{13}}\le O(1/\sqrt{N_{\rm c}})$.

\subsection{Resulting narrowness and number of tetraquarks:
``always two there are, \dots\ no less'' \cite{Y}}In summary, at
large $N_{\rm c}$, we expect (1) any flavour-exotic tetraquarks to
be formed pairwise and (2) all (compact) tetraquarks to be narrow,
with decay widths decreasing not slower than $1/N_{\rm c}^2$
\cite{TQ1,TQ2}.\\[.91ex]\emph{Acknowledgement.} D.M.~was supported
by the Austrian Science Fund (FWF), project P29028-N27.

\end{document}